\documentclass[12pt,preprint]{aastex}
\usepackage{pslatex}
\usepackage{graphicx}

\begin{document}

\title{The Turbulent Heating Rate in
Strong  MHD Turbulence with Nonzero Cross Helicity}
\author{Benjamin D. G. Chandran}
\email{benjamin.chandran@unh.edu} 
\affil{Space Science Center and
Department of Physics, University of New Hampshire}
\author{Eliot Quataert}
\email{eliot@astro.berkeley.edu}
\affil{Department of Astronomy, University of California, Berkeley}
\author{Gregory G. Howes}
\email{gregory-howes@uiowa.edu}
\affil{Department of Physics \& Astronomy, University of Iowa}
\author{Joseph V. Hollweg}
\email{joe.hollweg@unh.edu} 
\affil{Space Science Center and
Department of Physics, University of New Hampshire}
\author{William Dorland}
\email{bdorland@umd.edu}
\affil{Department of Physics, University
 of Maryland}

\begin{abstract}
  Different results for the cascade power~$\epsilon$ in strong,
  incompressible MHD turbulence with nonzero cross helicity appear in
  the literature. In this paper, we discuss the conditions under which
  these different results are valid. Our conclusions can be expressed
  in terms of the density~$\rho$, the rms amplitudes $z^+$ and $z^-$
  of Alfv\'enic fluctuations propagating parallel and anti-parallel to
  the background magnetic field ${\bf B}_0$, and the correlation
  length (outer scale) measured perpendicular to~${\bf B}_0$, denoted
  $L_\perp$.  We argue that if $z^+ \gg z^-$ and if the $z^-$
  fluctuations are sustained by the reflection of $z^+$ fluctuations
  in a strong background magnetic field, then $\epsilon \sim \rho
  (z^+)^2 z^-/L_\perp$ as in previous studies by Hossain, Matthaeus,
  Dmitruk, Lithwick, Goldreich, Sridhar, and others. On the other
  hand, if the minority wave type ($z^-$) is sustained by some form of
  forcing that is uncorrelated with or only weakly correlated with
  the~$z^+$ fluctuations, then $\epsilon$~can be much less than $\rho
  (z^+)^2 z^-/L_\perp$, as in previous studies by Dobrowolny,
  Lazarian, Chandran and others.  The mechanism for generating the
  minority wave type strongly affects the cascade power because it
  controls the coherence time for interactions between oppositely
  directed wave packets at the outer scale.
\end{abstract}

\maketitle

\section{Introduction}
\label{sec:intro} 

Turbulence plays an important role in numerous astrophysical plasmas,
from the solar corona and solar wind (Coleman 1968; Belcher \& Davis
1971; Tu \& Marsch 1995; Goldstein, Matthaeus, \& Roberts 1995) to
star-forming molecular clouds and clusters of galaxies (Boldyrev,
Nordlund, \& Padoan 2002; Narayan \& Medvedev 2001; Cho \&
Lazarian~2003; Elmegreen \& Scalo 2004).  In general, astrophysical
turbulence is compressible, and the compressive component of the
turbulence may play an important role in processes such as ion heating
and the scattering of energetic particles (Li \& Habbal 2001; Yan \&
Lazarian 2004; Chandran 2005, 2008a).  However, in this paper we focus
on the non-compressive component of the turbulence, which generally
dominates the energy at scales smaller than the collisional mean free
path, since (non-compressive) Alfv\'en waves experience negligible
linear collisionless damping at wavelengths~$\lambda$ much larger than
the proton gyroradius~$r_{i}$ and frequencies~$\omega$ much smaller
than the proton cyclotron frequency~$\Omega_i$ (Barnes 1966).

For $\lambda \gg r_{\rm i}$ and $\omega \ll \Omega_i$, the
non-compressive component of the turbulence can be modeled using
incompressible magnetohydrodynamics (MHD), even in  a low-collisionality
plasma. [See  Howes et~al~(2008) and Schekochihin
et~al~(2009) for recent discussions of this point.]
In incompressible MHD, the fluctuating velocity and magnetic fields
can be decomposed into fluctuations propagating parallel and
anti-parallel to the background magnetic field~${\bf B}_0$, denoted
$z^+$ and $z^-$ respectively. When the energies in $z^+$ and $z^-$
fluctuations differ, the turbulence has nonzero cross
helicity\footnote{The cross helicity is defined as $\int d^3\! x\:
  {\bf v} \cdot {\bf B}$, where ${\bf v}$ is the velocity and {\bf B}
  is the magnetic field.} and is referred to as ``imbalanced.''

Turbulence in the solar corona and inner solar wind is imbalanced,
with an excess of Alfv\'enic fluctuations propagating away from the
Sun (Roberts et~al~1987; Grappin et~al~1990; Bavassano et~al~2000a,
2000b; Bruno \& Carbone~2005). In order to model the heating of the
solar corona and solar wind, it is of interest to determine the
turbulent heating rate or cascade power in imbalanced MHD
turbulence. Different expressions for this cascade power have appeared
in the literature (see, e.g., Dobrowolny, Mangeney, \& Veltri~1980;
Hossain et~al 1995; Lithwick, Goldreich, \& Sridhar 2007; Beresnyak \&
Lazarian 2008; Chandran 2008b). Our goal in this paper is to clarify
the conditions under which these different expressions are valid.  We
focus on the difference between situations in which the minority
Alfv\'en wave type (which we take to be~$z^-$) is sustained by the
reflection of $z^+$ fluctuations and situations in which the $z^-$
fluctuations are sustained by some form of forcing.  Although a number
of studies have addressed the effects of cross helicity on the
inertial-range power spectra of $z^+$ and $z^-$ fluctuations (e.g.,
Grappin, Pouquet, \& L\'eorat 1983; Lithwick et~al 2007; Beresnyak \&
Lazarian 2008, 2009; Chandran 2008b; Perez \& Boldyrev~2009; Podesta
\& Bhattacharjee~2009), we do not address the inertial-range power
spectra in this paper.

\section{Theoretical Framework}
\label{sec:bg} 

We consider Alfv\'enic (i.e., transverse and non-compressive)
fluctuations in a stationary, inhomogeneous medium, with
background density $\rho$ and background magnetic and velocity
fields~${\bf B}_0$ and ${\bf u} = u {\bf B}_0/B_0$.  The velocity
and magnetic-field fluctuations~$\delta {\bf v}$ and $\delta {\bf
  B}$ can be written in terms of the Elsasser variables 
\begin{equation}
{\bf   z^\pm} = \delta {\bf v} \mp \frac{\delta {\bf B}}{\sqrt{4\pi\rho}}.
\label{eq:defzpm} 
\end{equation} 
The equations for ${\bf z}^\pm$ are (Velli 1993, Verdini \& Velli
2007)
\begin{equation}
\frac{\partial {\bf z}^\pm}{\partial t}
+ \left({\bf u} \pm {\bf v}_{\rm A}\right) \cdot \nabla{\bf z}^\pm
+ {\bf z}^\mp \cdot \nabla ({\bf u} \mp {\bf v}_{\rm A})
\mp \frac{1}{2} \left({\bf z}^\pm - {\bf z}^\mp\right)
\left(\nabla \cdot {\bf v}_{\rm A}  \mp \frac{1}{2} \nabla \cdot {\bf u}\right)
= - {\bf z}^\mp \cdot \nabla {\bf z}^\pm - \frac{\nabla p_{\rm tot}}{\rho},
\label{eq:els1} 
\end{equation} 
where ${\bf v}_{\rm A} = {\bf B}_0/\sqrt{4\pi \rho}$ and $p_{\rm tot}
= p + B^2/8\pi$ is the sum of the thermal and magnetic pressure.  The
second term on the left-hand side of equation~(\ref{eq:els1})
represents wave propagation at velocity $\pm {\bf v}_{\rm A}$ in the
frame of the plasma. The third and fourth terms on the left-hand side
give rise to wave reflection.  The right-hand side of
equation~(\ref{eq:els1}) contains the nonlinear terms. Nonlinear
interactions arise only when oppositely propagating wave packets
interact (Iroshnikov 1963; Kraichnan 1965).  In this paper, we focus
on turbulent fluctuations at the ``outer scale'' or energy-injection
scale.  We view these outer-scale fluctuations as a collection of
$z^+$ and $z^-$ wave packets with correlations lengths~$L_\perp$ and
$L_\parallel^\pm$ perpendicular and parallel to~${\bf B}_0$, where for
simplicity we have taken $L_\perp$ to be the same for both the $z^+$
and $z^-$ fluctuations.

The shearing of a $z^\pm$ wave packet at the outer scale is
dominated by interactions with $z^\mp$ wave packets at the outer
scale. Since we take the fluctuations to be Alfv\'enic in nature,
${\bf z}^\pm$ is perpendicular to~${\bf B}_0$, and the
magnitude of the ${\bf z}^\mp \cdot \nabla {\bf z}^\pm$ term in
equation~(\ref{eq:els1}) is~$\sim z_0^- z_0^+/L_\perp$,
where $z^\pm_0$ is the rms amplitude of the $z^\pm$ fluctuations
at the outer scale. The rate of shearing of
$z^\pm$ wave packets by $z^\mp$ wave packets at the outer scale
is thus
\begin{equation}
\omega_{\rm shear}^\mp \sim \frac{z_0^\mp}{L_\perp}.
\label{eq:shear} 
\end{equation} 
Nonlinear interactions between $z^+$ and $z^-$ fluctuations
can be thought of as ``collisions'' between oppositely directed
wave packets. If we consider some point~P moving with
a $z^\pm$ wave packet at velocity~$\pm v_{\rm A} \hat{z}$,
the time required for this point to pass through an outer-scale
$z^\mp$ wave packet is approximately 
\begin{equation}
t_{\rm coll}^\mp \sim \frac{L_\parallel^\mp}{v_{\rm A}}
\label{eq:tcol} 
\end{equation} 
(where we have ignored the factor of~$1/2$ that results from the
fact that the relative speed of point~P and the moving $z^\mp$ wave
packet is~$2v_{\rm A}$).  We define the quantity
\begin{equation}
\chi^\mp = \frac{z_0^\mp L_\parallel^\mp}{L_\perp v_{\rm A}} \sim 
\omega_{\rm shear}^\mp t_{\rm coll}^\mp.
\label{eq:defchi} 
\end{equation} 
If $\chi^\mp < 1$, then $\chi^\mp$ is approximately the fractional change in an
outer-scale $z^\pm$ wave packet caused by shearing by outer-scale
$z^\mp$ fluctuations during a single ``collision'' of duration~$t_{\rm
  coll}^\mp$. If $\chi^\mp \gtrsim 1$, then an outer-scale $z^\pm$ wave
packet undergoes a fractional change of order unity in a time~$\sim
L_\perp / z^\mp$, which is~$\lesssim t_{\rm coll}^\mp$.

We take $z^+$ to be the dominant wave type, and assume that
\begin{equation}
z^-_0 \ll z^+_0 \la v_A.
\label{eq:dominant} 
\end{equation} 
which implies that $\delta B \la B_0$.  Throughout this paper, we
focus on the strong-turbulence limit, in which 
\begin{equation}
 \chi^+ \ga 1.
\label{eq:chiplus1} 
\end{equation} 
We note that the conditions $\chi^+ \ga 1$ and $\delta B \ll B_0$ are
compatible if $L_\perp/ L_\parallel^+$ is sufficiently small.

\section{Forced Turbulence}
\label{sec:forced} 

If a magnetized plasma is physically stirred by a set of objects
distributed throughout its volume, and if the motions of the different
stirring objects are uncorrelated, then this stirring
will excite both $z^+$ and $z^-$ waves, with comparable power going
into each wave type. If this stirring is the primary mechanism for
exciting Alfv\'enic fluctuations, then the resulting turbulence will
be approximately balanced, with little cross helicity.  However,
volumetric forcing can play an important role in imbalanced turbulence
if, in addition to the forcing, a large flux of the dominant $z^+$
fluctuations is launched into the plasma from the plasma boundary.  If
the forcing is weak, then $z_0^-$ will be small, the cascade time of
the $z^+$ fluctuations will be large, and the $z^+$ fluctuations
injected at the plasma boundary will propagate a large distance into
the plasma before their energy cascades. The role of forcing is then
different for $z^+$ and $z^-$.  Within a certain distance of the
boundary (see section~\ref{sec:ex}), forcing makes only a small
contribution to the energy of the $z^+$ fluctuations, but is the
mechanism that sustains the~$z^-$ fluctuations. We note that if $z^+$
fluctuations satisfying equation~(\ref{eq:chiplus1}) are launched into
the plasma from the plasma boundary, and if $z^-$ energy is not
continually injected into the system throughout its volume, then $z^-$
energy will decay on the short time scale $ L_\perp/z_0^+$, leaving
the system in a maximally aligned state\footnote{When ${\bf z}^- =
  \delta {\bf v} + \delta {\bf B}/\sqrt{4\pi \rho} =0$, the state is
  called ``maximally aligned'' because $\delta {\bf v}$ and $\delta
  {\bf B}$ are everywhere anti-parallel.}  with vanishing cascade
power (Dobrowolny et~al~1980; Matthaeus \& Montgomery~1980; Pouquet,
Sulem, \& Meneguzzi~1988).

We model the forcing of $z^\pm$ fluctuations as a stochastic source
term~${\bf F}({\bf x},t)$ added to right-hand side of 
equation~(\ref{eq:els1}).  For simplicity, we take the spatial
variation of ${\bf F}$ in directions perpendicular to ${\bf B}_0$ to be
characterized by a single scale length (correlation length), which we
set equal to the perpendicular correlation length of the outer-scale
$z^+$ fluctuations,~$L_\perp$. We also take the spatial
variation of ${\bf F}$ along ${\bf B}_0$ to be characterized by a single
scale length, denoted $L_{\parallel, {\rm f}}$, which can differ
from~$L_{\parallel}^+$.  We assume that the (stochastic) time variation of ${\bf F}({\bf x},t)$
is characterized by the single time scale~$L_{\parallel, {\rm f}}/v_{\rm A}$.

\begin{figure}[h]
\includegraphics[width=4in]{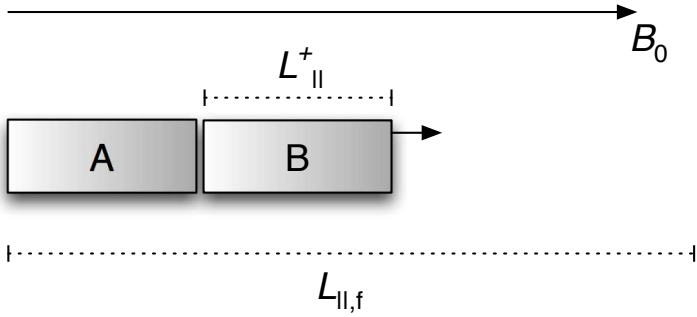}
\caption{\footnotesize A and B represent two outer-scale $z^+$ wave
  packets propagating along the background magnetic field~${\bf
    B}_0$. In this pictorial example, the parallel correlation length
  of the forcing term, $L_{\parallel, {\rm f}}$ is taken to be longer
  than the parallel correlation length of the outer-scale $z^+$ wave
  packets, $L_\parallel^+$.
\label{fig:f1}}
\end{figure}

The properties of outer-scale $z^-$ wave packets are determined by
both (1) the way in which the $z^-$ fluctuations are sheared by $z^+$
fluctuations and (2) the source term. We discuss these two mechanisms
in turn and their consequences for the correlation time of the
shearing that is applied by $z^-$ wave packets and experienced by a single,
moving $z^+$ wave packet. Because of equation~(\ref{eq:chiplus1}), $z^-$
fluctuations can only propagate a distance~$\lesssim L_\parallel^+$
before their energy cascades to smaller scales. As a result, if we
consider the outer-scale $z^+$ wave packet labeled ``A'' in
Figure~\ref{fig:f1}, the outer-scale $z^-$ fluctuations encountered by
wave~packet~A are sheared primarily by wave~packet~A and the
outer-scale wave packet immediately ``ahead'' of wave packet~A, which
is labeled wave~packet~B in Figure~\ref{fig:f1}.  Since wave packet~B
remains in front of wave packet~A as the two wave packets propagate
along the magnetic field, the shearing experienced by the outer-scale
$z^-$ fluctuations that are encountered by wave packet~A remains
coherent throughout the cascade time of wave packet~A.  On the other hand, the
properties of the source term~${\bf F}$ in the region that contributes
to the $z^-$ fluctuations encountered by wave packet~A decorrelates
over a time scale $\sim L_{\parallel, {\rm f}}/v_{\rm A}$, because of
both the space and time dependence of~${\bf F}$. Thus, the coherence
time for the shearing experienced by the outer-scale $z^+$ wave
packets is\footnote{Equation~(\ref{eq:tcor}) is valid unless
 $L_{\parallel,{\rm f}}/v_{\rm A}$ exceeds the $z^+$ cascade
time~$t_{\rm casc}^+$, in which case  $t_{\rm cor}^- \sim t_{\rm casc}^+$,
and  the cascade time and cascade
power are given by equations~(\ref{eq:epsm1}), (\ref{eq:tcascpl2}),
and  (\ref{eq:epsp3}).}
\begin{equation}
t^-_{\rm cor} \sim \frac{L_{\parallel, {\rm f}}}{v_{\rm A}}.
\label{eq:tcor} 
\end{equation} 
We note that if $L_{\parallel, {\rm f}} \gg L_{\parallel}^+$, then
because of equation~(\ref{eq:chiplus1}) the shearing experienced by
outer-scale $z^-$ fluctuations will cause the $z^-$ fluctuations to
vary strongly over a distance~$L_\parallel^+$ along the field, causing
$L_\parallel^-$ to become comparable to~$L_\parallel^+$. However, in
this case, the shearing experienced by the $z^+$ fluctuations remains
coherent for a time $L_{\parallel,{\rm f}}/v_{\rm A}$ rather than the
collision time scale $t_{\rm coll}^- \sim L_\parallel^-/v_{\rm A}$,
for the reasons given above.

We define the quantity
\begin{equation}
\hat{\chi}^- = \frac{z_0^- L_{\parallel, {\rm f}}}{v_{\rm A} L_\perp}.
\label{eq:defchihat} 
\end{equation} 
We first consider the case in which
\begin{equation}
\hat{\chi}^- < 1.
\label{eq:chihatm} 
\end{equation} 
When equation~(\ref{eq:chihatm})  is satisfied,
$\hat{\chi}^-$ is approximately the fractional change in an outer-scale
$z^+$ wave~packet during a time~$t_{\rm cor}^-$.  During successive
time intervals of duration $ t_{\rm cor}^-$, the incremental changes in
the $z^+$ wave packet that are caused by the shearing of~$z^-$ fluctuations
add like a random walk. Thus, $(\hat{\chi}^-)\,^{-2}$ time intervals of
duration~$t_{\rm cor}^-$ are required in order for the incremental
changes in the $z^+$ wave packet to accumulate to an order-unity
fractional change.  The cascade time for outer-scale $z^+$ wave
packets is then given by
\begin{equation}
t_{\rm casc}^+ \sim 
\frac{L_\perp^2 v_{\rm A}}{(z_0^-)^2 L_{\parallel,{\rm f}}}
\label{eq:deftcasc1} 
\end{equation} 
The cascade power in $z^+$ fluctuations, $\epsilon^+ \sim \rho
(z_0^+)^2/t_{\rm casc}^+$, is then (Beresnyak \& Lazarian~2008,
Chandran~2008b)
\begin{equation}
\epsilon^+ \sim \frac{\rho (z_0^+)^2 (z_0^-)^2 
L_{\parallel, f}}{v_{\rm A} L_\perp^2}.
\label{eq:epsp1} 
\end{equation} 
If we set $L_\perp = L_{\parallel,{\rm f}}$, then
equation~(\ref{eq:epsp1}) reduces to the result of Dobrowolny
et~al~(1980) for isotropic turbulence.  In the case that $L_{\parallel, {\rm f}}
\sim L_\parallel^+$ and $\chi^+ \sim 1$,
we can rewrite equation~(\ref{eq:epsp1})
as (Chandran~2008b)
\begin{equation}
\epsilon^+ \sim \frac{\rho z_0^+ (z_0^-)^2 }{L_\perp}.
\label{eq:epsp2} 
\end{equation} 
Because of equation~(\ref{eq:chiplus1}), the cascade time for $z^-$
fluctuations is simply $t_{\rm casc}^- \sim L_\perp/z_0^+$, and the
cascade power in $z^-$ fluctuations is 
\begin{equation}
\epsilon^- \sim \frac{\rho z_0^+ (z_0^-)^2}{L_\perp},
\label{eq:epsm1} 
\end{equation} 
comparable to the value of~$\epsilon^+$ in
equation~(\ref{eq:epsp2}).\hspace{0.1cm}\footnote{In the
  weak-turbulence limit $\chi^- \ll \chi^+ \ll 1$, both $\epsilon^+$
  and $\epsilon^-$ scale in the same way with $z_0^\pm$, $L_\perp$,
  and $L_\parallel$ [in particular, $\epsilon^\pm \propto \rho
  (z_0^+)^2 (z_0^-)^2 L_\parallel/(v_A L_\perp^2)$ (Lithwick \&
  Goldreich 2003)].  However, there is an additional dependence of
  $\epsilon^\pm$ on the inertial-range spectral indices, so that the
  fluctuation type ($z^+$ or $z^-$) with the steeper inertial-range
  power spectrum has the larger cascade power. It was shown
  analytically by Galtier et~al~(2000) and Lithwick \&
  Goldreich~(2003) that this dependence becomes very strong as the
  spectral indices approach the limiting values at which the
  assumption of local interactions breaks down, so that the ratio
  $\epsilon^+/\epsilon^-$ can take on any value. In the strong
  turbulence limit $\chi^- \ll \chi^+ \sim 1$, $\epsilon^+$ and
  $\epsilon^-$ also scale in the same way with $z_0^\pm$ and
  $L_\perp$ when $L_{\parallel, {\rm f}}\sim L_{\parallel}^+$
and $\hat{\chi}^- < 1$, as shown in equations~(\ref{eq:epsp2}) and
  (\ref{eq:epsm1}).  However, if $\epsilon^+$ and $\epsilon^-$ depend
  on the inertial-range spectral indices in the strong turbulence case
  as they do in the weak-turbulence limit, then additional
  coefficients (not necessarily of order unity) need to be included in
  equations~(\ref{eq:epsp2}) and (\ref{eq:epsm1}). The presence of
  such coefficients would allow for the possibility of forced, steady-state,
  strongly turbulent flows in which $\epsilon^+/\epsilon^-$ differs
  significantly from unity, even when the assumptions leading to
  equations~(\ref{eq:epsp2}) and (\ref{eq:epsm1}) are valid (i.e.,
  $\chi^+ \simeq 1$, $\hat{\chi}^- < 1$, and $L_{\parallel, {\rm f}}
  \sim L_\parallel^+$).}

If,  instead of equation~(\ref{eq:chihatm}), 
\begin{equation}
\hat{\chi}^- >1 
\label{eq:lim2} 
\end{equation} 
then the shearing experienced by outer-scale $z^+$ wave packets
remains coherent throughout their cascade time,
\begin{equation}
t_{\rm casc}^+ \sim \frac{L_\perp}{z_0^-},
\label{eq:tcascpl2} 
\end{equation} 
and
\begin{equation}
\epsilon^+ \sim \frac{\rho (z_0^+)^2 z_0^-}{L_\perp},
\label{eq:epsp3} 
\end{equation} 
while $\epsilon^-$ is still given by equation~(\ref{eq:epsm1}).
Turbulent heating rates of this form were previously obtained
by Hossain et~al (1995) on the basis of 
numerical simulations of decaying MHD turbulence,
and by Lithwick et~al~(2007) on the basis
of theoretical arguments.

\subsection{An Example of Forced, Imbalanced MHD Turbulence}
\label{sec:ex} 

In this section, we consider a finite plasma that is stirred by a set of
objects located throughout its volume.
We take the dimension and spacing of these objects perpendicular to the magnetic field
to be~$L_\perp$, and we take the objects to be spaced by a distance $L_\parallel^+$
along the magnetic field. We take the correlation time of the forcing
to be~$L_\parallel^+/v_{\rm A}$.
In the absence of wave launching at the boundary,
these stirring objects set up balanced MHD turbulence, in which the rms amplitudes
of the $z^+$ and $z^-$ fluctuations are both equal to
a quantity that we denote~$\delta z$. We define
\begin{equation}
\chi_0 = \frac{L_\parallel^+ \delta z }{L_\perp v_{\rm A}}
\label{eq:defchi0} 
\end{equation} 
and assume that
\begin{equation}
\chi_0\ll 1.
\label{eq:valchi0} 
\end{equation}
Thus, in the absence of wave launching, the forcing would set up 
weak turbulence at the outer scale, with an outer-scale cascade time
of $(\chi_0)^{-2} \,L_\parallel^+/v_{\rm A}$ 
and a cascade power of
\begin{equation}
Q_0 \sim \frac{\rho (\delta z)^4 L_\parallel^+}{L_\perp^2 v_{\rm A}}.
\label{eq:defQ0} 
\end{equation} 

We now assume that $z^+$ waves are launched into the plasma from one
its boundaries (the ``launching boundary'') with correlation lengths
$L_\perp$ and $L_\parallel^+$ perpendicular and parallel to the
magnetic field.  We consider the region that is sufficiently close to
the launching boundary that the amplitudes of the $z^+$ fluctuations
have not decayed very much from their values right next to the
boundary.  (The size of this region is estimated below.) We assume
that within this region,
\begin{equation}
\chi^+ \simeq 1,
\label{eq:chipex} 
\end{equation} 
implying that $z_0^+ \gg \delta z$.  We also assume that the presence
of the additional $z^+$ waves does not substantially alter the rate at which energy
is injected into either the $z^+$ of $z^-$ fluctuations.\footnote{This
  assumption is plausible for the following reason. The net power per
  unit volume going into the $z^\pm$ fluctuations, denoted $Q^\pm$, is
  approximately $\langle {\bf F}\cdot{\bf z}^\pm \rangle$, where
  $\langle \dots \rangle$ represents an ensemble average.  The value
  of ${\bf z}^\pm$ can be thought of as consisting of two components,
  ${\bf z}^\pm = {\bf z}_1^\pm + {\bf z}_2^\pm$, where ${\bf z}_1^\pm$
  is correlated with ${\bf F}$, and ${\bf z}_2^\pm$ is not. A rough
  estimate of $Q^\pm$ is then $F_{\rm rms} z_{1, {\rm rms}}^\pm$,
  where $F_{\rm rms}$ and $z_{1,{\rm rms}}^\pm$ are the rms
  amplitudes of ${\bf F}$ and ${\bf z}_1^\pm$. Because the
  correlation time of ${\bf F}$ in our example is
  $L_\parallel^+/v_{\rm A}$, which is comparable to or shorter than
  the cascade time of each fluctuation type, $z_{1,{\rm rms}}^\pm \sim
  F_{\rm rms} L_\parallel^+/v_{\rm A}$ for both wave types, so that
  $Q^+ \sim Q^-$.}
Since the cascade time for the $z^-$ fluctuations is short, the rate at
which energy is injected into $z^-$ fluctuations will be in balance with
the rate at which $z^-$ energy cascades to small scales. Thus,
\begin{equation}
Q_0 \sim \frac{\rho (z_0^-)^2 z_0^+}{L_\perp}.
\label{eq:zmex1} 
\end{equation} 
Equations~(\ref{eq:defQ0}) through (\ref{eq:zmex1}) then lead to the relation
\begin{equation}
z_0^- \sim \frac{(\delta z)^2 }{z_0^+},
\label{eq:zmex2} 
\end{equation} 
which illustrates how the enhanced level of $z^+$ fluctuations reduces
$z_0^-$ relative to its value ($\delta z$) in the absence of wave
launching from the boundary.

Determining the rate at which $z^+$ energy decays in this example is subtle.
Equations~(\ref{eq:epsp1}) and (\ref{eq:zmex2}) 
imply that 
\begin{equation}
\epsilon^+ \sim \frac{\rho (\delta z)^4 L_\parallel^+}{L_\perp^2 v_{\rm A}},
\label{eq:epspex} 
\end{equation} 
which is comparable to the rate at which energy is injected by the
forcing term.  The question of whether the $z^+$ energy increases or
decreases with increasing distance from the launching boundary thus
depends upon the undetermined factors of order unity in
equations~(\ref{eq:defQ0}) and (\ref{eq:epspex}). To proceed, we
conjecture that the $z^+$ and $z^-$ power spectra are ``pinned''
(i.e., equal) at the dissipation scale (Grappin et~al 1983; Lithwick
\& Goldreich 2003 --- but see Perez \& Boldyrev~2009 and Beresnyak \&
Lazarian~2009), and that there is a dimensionless prefactor in the
expression for $\epsilon^+$ (or $\epsilon^-$) that increases as the
$z^+$ (or $z^-$) inertial-range power spectrum becomes steeper, as in
the case of imbalanced weak MHD turbulence (Galtier et~al 2000;
Lithwick \& Goldreich 2003). This implies that in the imbalanced case
in which $z_0^+ \gg z_0^-$, the value of $\epsilon^+$ will be somewhat
larger than both $\epsilon^-$ and $Q_0$, so that the $z^+$ energy will
decay in time, when viewed from a frame of reference moving with the
$z^+$ fluctuations at velocity $v_{\rm A} \hat{z}$.  The time
scale~$t_{\rm casc}^+$ for this decay is given approximately by
equation~(\ref{eq:deftcasc1}), with $L_{\parallel, {\rm f} }$ set
equal to~$L_\parallel^+$.  In the plasma frame, the $z^+$ energy then
decays over a distance of $L_{\rm decay}^+ \sim v_{\rm A} t_{\rm
  casc}^+$, or (since $\chi^+ \simeq 1$)
\begin{equation}
L_{\rm decay}^+ \sim L_\parallel^+ \chi_0^{-4}.
\label{eq:Ldecay} 
\end{equation} 
At distances $\lesssim L_{\rm decay}^+$ from the launching boundary, the
turbulence would then be highly imbalanced, with forcing sustaining
the $z^-$ fluctuations but making only a small contribution to the
energy in $z^+$ fluctuations. On the other hand, at distances $\gg
L_{\rm decay}^+$ from the launching boundary the turbulence would
become approximately balanced. If our conjecture above about
the relation between the cascade power and inertial-range spectral
index in strong imbalanced turbulence is incorrect, then other phenomenologies are possible.

\section{Reflection-Driven Turbulence in a Strong Background
Magnetic Field}
\label{sec:reflection} 

In this section, we assume that the $z^-$ wave packets are
generated by reflection of the dominant $z^+$ wave packets.  The key
difference from the last section is that the mechanism generating the
$z^-$ fluctuations is now strongly correlated with, and in fact
determined by, the properties of the $z^+$ fluctuations.
We begin by considering outer-scale $z^+$ fluctuations that satisfy
the ``critical-balance condition'' (Goldreich \& Sridhar 1995; Lithwick
et~al~2007),
\begin{equation}
\chi^+ \sim 1.
\label{eq:cb} 
\end{equation} 
We consider the case $\chi^+ \gg 1$ at the end of this section.
Part of our motivation for considering outer scales satisfying
equation~(\ref{eq:cb}) comes from the work of Cranmer \& van
Ballegooijen (2005), who modeled the launching of Alfv\'en waves
into the corona and solar wind from the Sun. They based their
analysis on observations of photospheric motions and a model of
wave propagation in the chromosphere, corona, and solar wind.
They found that for heliocentric distances between 2 and~10 solar
radii, the bulk of the wave power is at periods of $1-5$~minutes,
with wave amplitudes such that equation~(\ref{eq:cb}) is
approximately satisfied. 

\begin{figure}[h]
\includegraphics[width=4in]{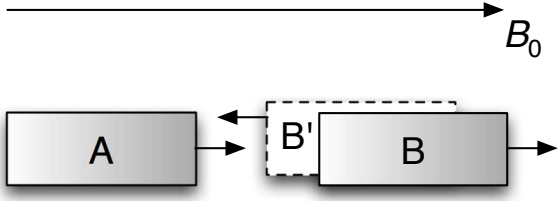}
\caption{\footnotesize In strong, reflection-driven turbulence, the
  properties of the $z^-$ wave packet ($\mbox{B}^\prime$) encountered
  by the $z^+$ wave packet on the left (A) are determined by the $z^+$
  wave packet on the right (B). The shearing of wave packet~A by wave
  packet~$\mbox{B}^\prime$ thus remains coherent over a time 
  comparable to the cascade time of wave packet~B.
\label{fig:f2}}
\end{figure}

Equation~(\ref{eq:cb}) implies that the outer-scale $z^-$ wave
packets only travel a distance~$\sim L_\parallel^+$ before their
energy cascades to smaller scales.  Thus, the outer-scale $z^-$
fluctuations present at an arbitrary point~P originated within a
distance~$\sim L_\parallel^+$ from~P. If we consider two outer-scale
$z^+$ wave packets A and~B of length~$L_\parallel^+$ and
width~$L_\perp$, with A right behind~B as in Figure~\ref{fig:f2}, then
the outer-scale $z^-$ wave packets encountered by wave-packet~A
(denoted ${\rm B}^\prime$ in Figure~\ref{fig:f2}) will primarily be
the $z^-$ field arising from the reflection of wave packet~B. If we
assume that the gradient scale lengths of the background medium are at
least as long as the ``damping length'' $v_{\rm A} t_{\rm casc}^+$ of
the $z^+$ wave packet, then the $z^-$ wave packets encountered by wave
packet~A will continue to have the same form and structure until wave
packet~B changes. The shearing experienced by wave packet~A is thus
coherent throughout its cascade time, rather than a random-walk-like
accumulation of uncorrelated, small-amplitude distortions.  The
cascade time of wave packet A is therefore simply~$\sim (\omega_{\rm
  shear}^-)^{-1}$, or
\begin{equation}
t_{\rm casc}^+ \sim \frac{L_\perp}{z_0^-},
\label{eq:tcasc_ref} 
\end{equation} 
which implies that
\begin{equation}
\epsilon^+ \sim \frac{\rho (z_0^+)^2 z_0^-}{L_\perp},
\label{eq:epsp_ref} 
\end{equation} 
as in the studies of Hossain et~al (1995) and
Lithwick et~al~(2007). 

Lithwick et~al~(2007) argued that in strong,
imbalanced, MHD turbulence, the shearing felt by the $z^+$ wave
packets throughout the inertial range is coherent over a time that
significantly exceeds the wave-packet ``collision time'' or crossing
time, independent of the mechanism that generates the outer-scale
fluctuations.  Our arguments differ from theirs in two ways. First,
our discussion is restricted to the outer-scale fluctuations. Second,
we argue that equation~(\ref{eq:tcasc_ref}) is valid when the minority
$z^-$ wave type is generated by wave reflection, but not when~$z^-$ is
generated by forcing that is only weakly correlated with the $z^+$
fluctuations (unless $\hat{\chi}^- > 1$, as described in 
section~\ref{sec:forced}).

For the waves launched into the solar corona by photospheric motions,
it is typically argued that $L_\perp \sim 10^9 \mbox{ cm}$ at the
coronal base, and that the Alfv\'en waves at this perpendicular scale
have a broad frequency spectrum, extending up to wave periods of many
hours (Dmitruk \& Matthaeus 2003, Cranmer \& van Ballegooijen 2005,
Verdini \& Velli 2007). Although Cranmer \& van Ballegooijen (2005)
find that the energy in outward Alfv\'en waves ($z^+$ in our
discussion) in the corona is dominated by waves with periods of $\sim
1-5$~minutes, wave reflection is more efficient for low-frequency
waves. Thus, if nonlinear wave-wave interactions were neglected, then
the inward wave packets would have a longer parallel correlation
length than the outward waves.

We now explore the effects of frequency-dependent reflection,
including the effects of nonlinear wave-wave interactions, by
considering the limiting case in which the source term for inward
($z^-$) waves is invariant along the magnetic field.  The parallel
correlation length of the $z^-$ fluctuations is given by the
arguments of Lithwick et~al~(2007), as follows.  We continue
to assume the validity of equation~(\ref{eq:cb}), which is motivated 
for the coronal case by the results of Cranmer \& van
Ballegooijen's (2005) model of wave launching and propagation.
Equation~(\ref{eq:cb}) implies that a $z^-$ wave packet can only
travel a distance of order~$L_\parallel^+$ before the wave packet
is strongly distorted.  Since the $z^+$ fluctuations decorrelate
over a distance~$L_\parallel^+$ along the magnetic field, the
inward waves acquire a parallel correlation length
of~$L_\parallel^+$ by interacting with the outward waves, even
though the parallel correlation length of the source term for the
inward waves is~$\gg L_\parallel^+$.

Nevertheless, the coherence time for the shearing experienced by an
outer-scale $z^+$~wave packet remains much longer than~$t_{\rm
  coll}^-$. This can again be seen with the aid of Figure~\ref{fig:f2}.
Now, the source term for the inward waves is invariant along the
magnetic field, but the properties of the $z^-$ wave packet (${\rm
  B}^\prime$) encountered by the $z^+$ wave packet~A are determined by
the way in which the $z^+$ wave packet~B shears the inward waves
generated by the uniform source term. Since wave packet A and B remain
adjacent as they propagate outward along the magnetic field, the $z^-$
wave field encountered by wave packet~A will remain approximately the
same until wave packet~B changes, i.e., for a time equal to the
cascade time of the outward waves. Since the shearing of the $z^+$
wave packets is coherent throughout their cascade time, their cascade
time is again given by equation~(\ref{eq:tcasc_ref}), and the cascade
power for~$z^+$ fluctuations is again given by
equation~(\ref{eq:epsp_ref}).

We now turn briefly to the case in which $\chi^+ \gg 1$, with the
$z^-$ fluctuations generated by wave reflection.  As in the case
$\chi^+ \sim 1$, the shearing experienced by outer-scale $z^+$ wave
packets remains coherent throughout their cascade time, and the
cascade power in $z^+$ fluctuations is again given by
equation~(\ref{eq:epsp_ref}), as in the studies by Matthaeus
et~al~(1999) and Dmitruk \& Matthaeus (2003).

Although we have focused on wave reflection in this section, similar
arguments would apply to a compressible plasma in which
the minority waves ($z^-$) are generated by the
parametric decay of $z^+$ waves (Galeev \&
Oraevskii 1963; Sagdeev \& Galeev 1969; Vi\~nas \& Goldstein 1991;
Malara et~al~2000, 2001; Del Zanna et~al 2001). In this case, the
$z^-$ waves encountered by the $z^+$ wave packet~A in
Figure~\ref{fig:f2} will again be determined primarily by the
properties of the $z^+$ wave packet~B, which is immediately ahead of
wave packet~A, and the shearing experienced by wave packet~A will
remain coherent throughout the cascade time of wave packet~A.

\section{Application to the Solar Corona and Solar Wind}

In the solar corona, the dominant source of low-frequency,
Sunward-propagating, Alfv\'en waves is likely wave reflection
(Matthaeus et~al~1999; Dmitruk et~al~2001, 2002).  The anti-Sunward
waves launched by photospheric motions have sufficiently large
amplitudes and low frequencies that $\chi^+ \gtrsim 1$ within the
corona (Cranmer \& van Ballegooijen 2005; Cranmer, van Ballegooijen,
\& Edgar 2007; Verdini \& Velli 2007). If these anti-Sunward waves
have enough time to develop a turbulent power spectrum extending to
the (perpendicular) dissipation scale before they propagate out of the
corona to larger heliocentric distances, then the cascade power and
turbulent heating rate in the corona are given by
equation~(\ref{eq:epsp_ref}), as in the studies of Matthaeus
et~al~(1999), Dmitruk \& Matthaeus (2003), Cranmer \& van Ballegooijen
(2005), and Verdini \& Velli (2007). The application of
equations~(\ref{eq:epsp1}) and (\ref{eq:epsp2}) to the corona by
Chandran~(2008b) was incorrect.  The use of the correct equation in
models of the coronal is critical, because the energy of anti-Sunward
waves dramatically exceeds the energy of Sunward waves in the
corona (Cranmer \& van Ballegooijen~2005; Verdini \& Velli 2007). When
the correct expressions for $\epsilon^+$ and $\epsilon^-$ are used in
models of the turbulent heating of open-magnetic-field-line regions in
the solar corona, the heating rate is energetically sufficient to
explain the acceleration of the fast solar wind for plausible
assumptions about the amplitude and frequency spectrum of the waves
launched from the Sun (Matthaeus et~al~1999; Dmitruk et~al 2002;
Cranmer \& van Ballegooijen 2005; Verdini \& Velli 2007). On the other
hand, if equations~(\ref{eq:epsp1}) and/or (\ref{eq:epsp2}) were used,
then the heating rate in these models would be much too small to
explain the acceleration of the fast wind, because $z_0^-/z_0^+$ is so
small in the corona. [See Cranmer \& van Ballegooijen~(2005) and
Verdini \& Velli~(2007).]

In the solar wind at low heliographic latitudes, the
Sunward-propagating Alfv\'en waves may be driven primarily by
instabilities resulting from shear in the solar-wind
velocity (Roberts et~al~1992; Breech et~al~2008). It is not entirely
clear how to model this process, but a reasonable first approximation
is to treat the generation of Sunward waves by velocity-shear
instabilities as a form of mechanical forcing that is only weakly
correlated with the anti-Sunward waves. If we denote the parallel
correlation length of this forcing as $L_{\parallel, {\rm f}}$, and if
$\hat{\chi}^- < 1$, then the cascade power and turbulent heating rate
are given by equation~(\ref{eq:epsp1}), which is a factor of
$\hat{\chi}^-$ smaller than the value in
equation~(\ref{eq:epsp_ref}) (Beresnyak \& Lazarian~2008;
Chandran~2008b). If $L_{\parallel, {\rm f}} \sim L_\parallel^+$ and
$\chi^+ \sim 1$, then the cascade power and turbulent heating rate are
given by equation~(\ref{eq:epsp2}) (Chandran~2008b). The difference
between the different expressions for the turbulent heating rate in
reflection-driven turbulence and forced turbulence makes less of a
difference in models of the solar wind at heliocentric distances of
1~AU and beyond than in models of the corona, because $z_0^-/z_0^+$ is
not so small. Nevertheless, this difference may be significant when
comparing model temperature profiles with observations, as shown in
the recent study by Ng et~al~(2009).

\section{Conclusion}

In this paper, we have argued that the cascade power in strong,
imbalanced, incompressible MHD turbulence depends upon the mechanism
that sustains the minority Alfv\'en wave type~($z^-$) at the outer
scale. If the outer-scale $z^-$ wave packets are sustained by the
reflection of $z^+$ wave packets, then the shearing of the outer-scale
$z^+$ wave packets remains coherent throughout their cascade time, a
time that can greatly exceed the crossing time of two
counter-propagating outer-scale wave packets. On the other hand, if
the outer-scale $z^-$ wave packets are sustained by some form of
forcing that is only weakly correlated with the~$z^+$ fluctuations,
then the coherence time for the shearing of the outer-scale $z^+$ wave
packets is shorter than the cascade time, unless the parallel
correlation length of the forcing function is very long (i.e., unless
$\hat{\chi}^->1$).  The longer coherence time in reflection-driven
turbulence enhances the cascade power in reflection-driven turbulence
relative to forced turbulence with the same values of $z_0^\pm$,
$L_\perp$, and $L_\parallel^\pm$.

\acknowledgements We thank Marco Velli, Bill Matthaeus, Steve Cranmer,
and the referee Yoram Lithwick for helpful discussions and
comments. This work was supported in part by the the Center for
Integrated Computation and Analysis of Reconnection and Turbulence
(CICART) under DOE Grant DE-FG02-07-ER46372, by 
NSF-DOE Grant AST-0613622, by NSF Grant ATM-0851005,
and by NASA under Grants NNX07AP65G
and NNX08AH52G. E. Quataert was supported in part by
NSF-DOE Grant PHY-0812811 and by NSF Grant ATM-0752503.

\references

Barnes, A. 1966, Phys. Fluids, 9, 1483

Bavassano, B., Pietropaolo, E.,\& Bruno, R. 2000a,
J. Geophys. Res., 105, 12697

Bavassano, B., Pietropaolo, E., \& Bruno, R. 2000b,
J. Geophys. Res., 105, 15959

Belcher, J. W., \& Davis, L. 1971, J. Geophys. Res., 76, 3534

Beresnyak, A., \& Lazarian, A. 2008, ApJ, 682, 1070

Beresnyak, A., \& Lazarian, A. 2009, arXiv:0904.2574

Boldyrev, S., Nordlund, A., \& Padoan, P. 2002, Phys. Rev. Lett., 89,
031102

Breech, B., Matthaeus, W. H., Minnie, J., Bieber, J. W.,
 Oughton, S., Smith, C. W., \& Isenberg, P. A. 2008,
J. Geophys. Res., 113, A08105

Bruno, R., \& Carbone, V. 2005, Living Rev. Solar Phys., 2, 4
[Online article: cited Nov. 14, 2007, http://www.livingreviews.org/lrsp-2005-4]

Chandran, B. D. G. 2005, Phys. Rev. Lett., 95, 265004

Chandran, B. D. G. 2008a, Phys. Rev. Lett., 101, 235004

Chandran, B. D. G. 2008b, ApJ, 685, 646

Cho, J., \& Lazarian, A. 2003, MNRAS, 345, 325

Coleman, P. J. 1968, ApJ, 153, 371

Cranmer, S. R. \& van Ballegooijen, A. A. 2005, ApJS, 156, 265

Cranmer, S. R., van Ballegooijen, A. A., \& Edgar, R. J. 2007, ApJS, 171, 520

Del Zanna, L., Velli, M., \& Londrillo, P.  2001, A\&A,  367, 705

Dmitruk, P., Milano, L. J., \&
Matthaeus, W. H. 2001, ApJ, 548, 482

Dmitruk, P., Matthaeus, W. H., Milano, L. J., Oughton, S., Zank, G. P.,
\& Mullan, D. J. 2002, ApJ, 575, 571

Dmitruk, P., \& Matthaeus, W. H. 2003, ApJ, 597, 1097

Dobrowolny, M., Mangeney, A., Veltri, P. L. 1980, Phys. Rev. Lett., 35, 144

Galeev, A. A., \& Oraevskii, V. N. 1963, Sov. Phys. Dokl., 7, 988

Galtier, S., Nazarenko, S. V., Newell, A. C., \& Pouquet,
A. 2000, J. Plasma Phys., 63, 447

Goldreich, P., \& Sridhar, S. 1995, ApJ, 438, 763

Goldreich, P., \& Sridhar, S. 1997, ApJ, 485, 680

Goldstein, B. E., Smith, E. J., Balogh, A., Horbury, T. S., Goldstein,
M. L., \& Roberts, D. A. 1995, Geophys. Res. Lett., 22, 3393

Goldstein, M. L., Roberts, D. A., \& Matthaeus, W. H. 1995,
Ann. Rev. Astron. Astrophys., 33, 283

Grappin, R., Pouquet, A., \& L\'eorat, J. 1983, A\&A, 126, 51

Grappin, R., Mangeney, A., \& Marsch, E. 1990, J. Geophys. Res., 95,
8197

Hossain, M., Gray, P. C., Pontius, D. H., Matthaeus, W. H., \& Oughton, S.
1995, Phys. Fluids, 7, 2886

Howes, G. G., Cowley, S. C., Dorland, W., Hammett, G. W., Quataert, E.,
\& Schekochihin, A. A. 2008, J. Geophys. Res., 113, A05103

Iroshnikov, P. 1963, Astron. Zh. 40, 742

Kraichnan, R.  H. 1965, Phys. Fluids 8, 1385

Li, X., \& Habbal, S. R. 2001, J. Geophys. Res., 106, 10669

Lithwick, Y., \& Goldreich, P. 2003, ApJ, 582, 1220

Lithwick, Y., Goldreich, P., \& Sridhar, S. 2007, ApJ, 655, 269 

Malara, F., Primavera, L., \& Veltri, P. 2000,  Phys. Plasmas, 7, 2866

Malara, F., Primavera, L., \& Veltri, P., 2001,
in {\em Recent Insights into the Physics of the Sun and Heliosphere:
Highlights from SOHO and Other Space Missions}, eds. Brekke, P., Fleck, B., \& Gurman,
J.B., {\em Proceedings of the 24th General Assembly of the IAU}  (IAU Symposium 203, pp. 511-513), (San Francisco: Astronomical Society of the Pacific)

Matthaeus, W. H., \& Montgomery, D. 1980,  New York Acad. Sci., 357, 203

Matthaeus, W. H., Zank, G. P., Oughton, S., Mullan, D. J., \& Dmitruk, P. 1999,
ApJL, 523, L93

Narayan, R., \& Medvedev, M. 2001, ApJ, 562, 129

Ng, C. S., \& Bhattacharjee, A. 1997, Phys. Plasmas, 4, 605
	
Ng, C. S., Bhattacharjee, A., Isenberg, P. A., Munsi, D., \& Smith, C. W., Kolmogorov Versus Iroshnikov-Kraichnan Spectra: Consequences for Ion Heating in the Solar Wind,
in preparation.

Perez, J. C., \& Boldyrev, S. 2009, Phys. Rev. Lett., 102, 025003 (2009)

Podesta, J., \& Bhattacharjee, A., Phys. Rev. Lett., submitted

Pouquet, A., Sulem, P. L., \& Meneguzzi, M. 1988, Phys. Fluids, 31, 2635

Roberts, D. A., Goldstein, M. L., Klein, L. W.,
\&  Matthaeus, W. H. 1987, J. Geophys. Res., 92, 12023

Roberts, D. A., Goldstein, M. L., Matthaeus, W. H.,
\& Ghosh, S. 1992, J. Geophys. Res., 97, 17115

Sagdeev, R. Z., \& Galeev, A. A. 1969, {\em Nonlinear Plasma
Theory} (Benjamin: New York)

Schekochihin, A. A., Cowley, S. C., Dorland, W., Hammett, G. W., Howes, G. G., Quataert, E., \&  Tatsuno, T 2009, ApJS, accepted

Smith, C. W. 2003, {\em Solar Wind Ten: Proceedings of the Tenth
  International Solar Wind Conference}, AIP Conference Proceedings,
679, 413

Tu, C. Y., \& Marsch, E., 1995, Sp. Sci. Rev., 73, 1

Velli, M. 1993, A\&A, 270, 304

Verdini, A., \& Velli, M. 2007, ApJ, 662, 669

Vi\~nas, A.F. \& Goldstein, M.L. 1991,  J. Plasma Phys., 46, 129

Yan, H., \& Lazarian, A. 2004, ApJ, 614, 757

\end{document}